\documentstyle[12pt]{article}

\begin{document}
\hoffset = -1.4truecm
\voffset = -1.5truecm
\date{}
\title{\bf Isospin Symmetry Breaking Effects on the Properties of
Asymmetrical Nuclear Matter and $\beta$-Stable Matter}
\author{{\bf G. H. Bordbar}
\\
Department of Physics, Shiraz University,
Shiraz 71454, Iran,\\
Research Institute for Astronomy and Astrophysics of Maragha (RIAAM), Iran\\
and\\
Institute for Studies in Theoretical Physics and Mathematics (IPM),\\
Tehran, P. O. Box 19395-5531, Iran
}
\maketitle
%************************************************************************
\begin{abstract}
We have studied the influences of isospin symmetry breaking of
nucleon-nucleon interaction on the various properties of asymmetrical
nuclear matter and $\beta$-stable matter.
For asymmetrical nuclear matter, it is found that by including this
isospin symmetry breaking, the changes of bulk properties increase by
increasing both density and asymmetry parameter.
However, these effects on the total energy and equation of state of
$\beta$-stable matter are ignorable.
For asymmetrical nuclear matter, the validity of the empirical parabolic
law in the isospin symmetry breaking case is shown.
It is observed that the isospin symmetry
breaking of nucleon-nucleon interaction affects the $\beta$-equilibrium
conditions in $\beta$-stable matter.
\end{abstract}
\noindent {\bf Key words:} nucleon-nucleon interaction; isospin symmetry
breaking; nuclear matter; $\beta$-stable matter.
%***********************************************************************
\section{Introduction}
\label{intro}
In the calculation of the properties of nucleonic matter,
the nucleon-nucleon
potential has a crucial role. Recent models of two-nucleon potentials which
fit the nucleon-nucleon scattering data with high precision, contain terms
which break the isospin symmetry (Stoks et al., 1994; Machleidt and Slaus,
2001; Li and Machleidt, 1998). Isospin symmetry is
the invariance under any rotation in the isospin space. But, due to the
mass difference between neutral pion and charged pion
($\Delta m\simeq 4.6 MeV$), this symmetry is broken
(Stoks et al., 1994; Machleidt and Slaus, 2001; Li and Machleidt, 1998).
This means that for the new two-nucleon potentials, 
the neutral pion exchange is distinguished from charged pion exchange.
This distiction implies that the neutron-neutron ({\it nn}), neutron-proton
({\it np}) and proton-proton ({\it pp}) interactions are different and
therefore, these potentials are called isospin symmetry breaking ({\bf ISB})
potentials (Stoks et al., 1994).

Due to the distinction between the neutral pion and charged pion exchange,
for the potentials which describe the long range part of the nucleon-nucleon
interaction in terms of the one-pion exchange model, the one-pion exchange
term is one of the origins of {\bf ISB}
(Stoks et al., 1994; Machleidt and Slaus, 2001; Li and Machleidt, 1998).
For the {\bf ISB} potentials, in the
isospin $T=1$ states, the distinction between the {\it nn}, {\it np} and
{\it pp} potentials is necessary (Stoks et al., 1994).
Therefore, in the calculations
with these potentials, the difference between
the {\it nn}, {\it np} and {\it pp} wave functions should be considered.
In our calculations, we use the {\it Reid-93} potential
(Stoks et al., 1994).
This potential describes the longe range part of the nucleon-nucleon
interaction in terms of the one-pion exchange model accounting for the
mass difference between the neutral pion $\pi ^0$ and charged pions
$\pi^+$ and $\pi^-$. In the {\it Reid-93} potential, for the
non-one-pion exchange parts, additional {\bf ISB} terms have been included
in the $^1S_0$ state potential (Stoks et al., 1994).

In this work, we intent to calculate the various properties of asymmetrical
nuclear matter and $\beta$-stable matter in two different cases with
isospin symmetry ({\bf IS}) and with isospin symmetry breaking ({\bf ISB}).
By comparing the results of the {\bf IS} and
{\bf ISB} cases, we study the influence
of {\bf ISB} of nucleon-nucleon interaction
on these properties. For the {\bf ISB}
case, we use the {\it Reid-93} potential (Stoks et al., 1994)
in which the one-pion exchange
potentials for the {\it np} and {\it pp} ({\it nn}) interactions
as well as the {\it np} and {\it pp} ({\it nn}) $^1S_0$ potentials
are different. However, for the
{\bf IS} case, we replace the one-pion exchange and $^1S_0$ potentials
of {\it nn} and {\it pp} interactions with the corresponding {\it np}
interaction (Stoks et al., 1994).
In our calculations, we employ the formalism used in
our previous works (Bordbar and Modarres, 1997, 1998; Modarres and Bordbar,
1998; Bordbar and Riazi, 2001, 2002; Bordbar, 2002a,b)
in which we consider the
distiction between the partial waves with $M_T = +1$, $0$ and $-1$.
This method is a variational approach based on the cluster expansion of
the energy functional containing the state dependent correlation functions.
The convergence of the results of this approach has been also tested by
calculating the three-body cluster term (Bordbar and Modarres, 1997).
%****************************************************************************
\section{Isospin Symmetry Breaking Effects on the 
Asymmetrical Nuclear Matter Properties}
\label{Sec2}
The asymmetrical nuclear matter properties are important for studying
the supernova collapse and heavy-ion reaction (Lattimer and Prakash, 2000;
Li et al., 1997).
For asymmetrical nuclear matter, we define the asymmetry parameter $\beta$
as 
\begin{equation}
\beta=\frac{\rho_n -\rho_p}{\rho},
\end{equation}
where $\rho_n$ and $\rho_p$ are the number
densities of neutrons and protons
respectively, and $\rho = \rho_n+\rho_p$ is the total number density.
This implies that for symmetrical nuclear matter ($\rho_n=\rho_p$)
$\beta=0.0$ and for the pure neutron matter $\beta=1.0$.

In this section, we present our results for the various properties of
asymmetrical nuclear matter and then we investigate the effects of {\bf ISB}
on these properties.

The total energy of asymmetrical nuclear matter as a function of total
number density for various asymmetry parametrs
is given in Figure \ref{fig1}.
It is seen that for both
{\bf IS} and {\bf ISB} cases as
the asymmetry parameter increases, the saturation
point shifts to the lower densities.
However, for the large values of $\beta$,
the energy curve does not shows any minimum.
The changes of total energy due to the
inclusion of {\bf ISB} for nucleon-nucleon interaction are shown in Figure
\ref{fig2}.
This figure shows that the differences between our results in the
{\bf ISB} and {\bf IS} cases increase by increasing both density and
asymmetry parameter. In Table \ref{tab1}, our results for the differences
between the {\bf ISB} and {\bf IS} energies at $\beta = 0.0$ are compared
with those of MPM (Muther et al., 1999).
It can be seen that our results are in a good
agreement with the results of MPM for the {\it Nijm-II} potential.

The $\beta$-dependence of the total energy of asymmetrical nuclear matter
in the {\bf ISB} case is shown in Figure \ref{fig3}
for different values of density.
It can be seen that even in the
presence of {\bf ISB}, the asymmetrical nuclear
matter energy is nearly linear in $\beta^2$ and
therefore, the higher orders
effects of the asymmetry are very small. This confirms the validity
of the empirical parabolic law.

In Figure \ref{fig4}, we have shown the effects
of {\bf ISB} on the potential energy of
asymmetrical nuclear matter. It is seen that the potential energy
difference between the {\bf IS} and
{\bf ISB} cases increases by increasing asymmetry
parameter $\beta$. This is due to the fact that by increasing $\beta$,
the energy contribution of {\it nn} interaction increases.
Therefore, the difference
between the case in which we distinguish between the {\it nn} and {\it np}
interactions ({\bf ISB} case) and the case in which we replace the {\it nn}
interaction by the corresponding {\it np}
interaction ({\bf IS} case) increases
by increasing $\beta$. It is also seen that for each value of density
and asymmetry parameter, the potential energy shift due to the {\bf ISB}
is positive. This behaviour is reflecting the fact that in the partial
wave with isospin $T=1$, the $M_T =0$ potential is more attractive than
the corresponding $M_T =+1$ and $M_T=-1$ potentials (Bordbar, 2003).

The influence of {\bf ISB} on
the saturation point properties of asymmetrical
nuclear matter is shown in Table \ref{tab2}.
We see that the {\bf ISB} effects on the
saturation density, energy and incompressibility of asymmetrical nuclear
matter are generally small, but they are considerable for large $\beta$.

The nuclear matter asymmetry energy is of particular interest in
astrophysics, especially for studying the supernova mechanism
(Lattimer and Prakash, 2000).
The asymmetry energy can be approximated as the difference between binding energies
of pure neutron matter ($\beta=1.0$) and symmetrical nuclear matter
($\beta=0.0$).
The validity of using this approximation (called parabolic approximation)
in calculating the asymmetry energy was shown in Figure \ref{fig3}.
Our results for the asymmetry energy calculated for both
{\bf IS} and {\bf ISB} cases are presented in Figure \ref{fig5}.
It can be seen that the changes of asymmetry energy due to the inclusion of
{\bf ISB} increase by inceasing density.
%****************************************************************************
\section{Isospin Symmetry Breaking Effects on the 
$\beta$-Stable Matter Properties}
\label{Sec3}
The properties of $\beta$-stable matter are of special importance in
astrophysics. The equation of state of $\beta$-stable matter is very
important for investigating the neutron stars,
particularly in their stability (Akmal et al., 1998; Heiselberg and
Pandharipande, 2000).
The $\beta$-stable matter is a composition of nucleons and
leptons which is electrically neutral and in $\beta$-equilibrium.
The $\beta$-equilibrium means that there is a balance between the competing
processes of electron capture $e + p\rightarrow n +\nu$ and $\beta$-decay
$n\rightarrow e + p + \bar\nu$. This equilibrium conditions determine
the proton fraction ($x = \rho_p/\rho$) in $\beta$-stable matter.
The proton fraction is important for studing the cooling of a neutron star
(Prakash, 1994).

In Figure \ref{fig6}, we have plotted our results for the total energy of
$\beta$-stable matter versus density in the {\bf IS} and {\bf ISB} cases.
It is seen that by considering {\bf ISB} for the nucleon-nucleon interaction,
the change of $\beta$-stable matter energy is small. For example, at
density $\rho =1.0 fm^{-3}$,
the energy correction due to the {\bf ISB} effects
is about $1.6\%$.
Therefore, the effects of {\bf ISB} on
the total energy of $\beta$-stable matter
are negligible.

The calculated energy contribution of leptons and proton fraction in
$\beta$-stable matter are shown in Figures \ref{fig7} and \ref{fig8}.
It is found that these properties of $\beta$-stable matter are varied
by the inclusion of {\bf ISB} of nucleon-nucleon interaction.
This shows that the {\bf ISB} of nucleon-nucleon interaction affects the
conditions of $\beta$-equilibrium.

The equation of state of $\beta$-stable matter calculated for both {\bf IS}
and {\bf ISB} cases are shown in Figure \ref{fig9}.
It can be seen that the influences
of {\bf ISB} on the equation of state of
$\beta$-stable matter are ignorable.
%*****************************************************************************
\section{Summary and Conclusion}
\label{Sec4}
The two-nucleon potential has a crucial role in the nuclear many-body
calculations. By comparing the results of two different
cases of isospin symmetry
({\bf IS}) and isospin symmetry breaking ({\bf ISB}) for the nucleon-nucleon
interaction, we have evaluated the effects of {\bf ISB} on the different
properties of asymmetrical nuclear matter and $\beta$-stable matter.
It was seen that the changes of total energy, potential energy and
asymmetry energy of asymmetrical nuclear matter due to the inclusion
of {\bf ISB} for the nucleon-nucleon interaction increase by increasing
both density and asymmetry parameter.
It was shown that the empirical parabolic law is valid, even by
considering {\bf ISB} for the nucleon-nucleon interaction.
It was indicated that our results for the total
energy and the equation of state of $\beta$-stable matter in the {\bf IS}
and {\bf ISB} cases are nearly identical. For $\beta$-stable matter, it was
shown that the {\bf ISB} affects
the energy contribution of leptons and proton
fraction.
%*****************************************************************************
\section{Acknowledgment}
Financial support from Shiraz University research council and IPM is
gratefully acknowledged.
%%%%%%%%%%%%%%%%%%%%%%%%%%%%%%%%%%%%%%%%%%%%%%%%%%%%%%%%%%%%%%%%%%%%%%%%%%%

%********************************************************************
\newpage
\begin{figure}
\caption{ The total energy of asymmetrical nuclear matter versus density
for $\beta =$0.0, 0.33, 0.66 and 1.0 in the {\bf IS} (dashed curves) and
{\bf ISB} (full curves) cases.
}
\label{fig1}
\end{figure}
%********************************************************************
\begin{figure}
\caption{The difference between total energy of {\bf ISB} case ($E_{ISB}$)
and total energy of {\bf IS} case ($E_{IS}$)
versus asymmetry parameter $\beta$
for densities $\rho =$0.15 (full curve), 0.3 (dashed curve),
0.6 (dotted curve) $fm^{-3}$.
}
\label{fig2}
\end{figure}
%********************************************************************
\begin{figure}
\caption{
The total energy of asymmetrical nuclear matter in the {\bf ISB} case
versus $\beta^2$ for $\rho =$0.3 (full curve) and 0.5
(dotted curve) $fm^{-3}$.
}
\label{fig3}
\end{figure}
%%%%%%%%%%%%%%%%%%%%%%%%%%%%%%%%%%%%%%%%%%%%%%%%%%%%%%%%%%%%%%%%%%%
\begin{figure}
\caption{As Figure 1, but for the potential energy of asymmetrical
nuclear matter.
}
\label{fig4}
\end{figure}
%%%%%%%%%%%%%%%%%%%%%%%%%%%%%%%%%%%%%%%%%%%%%%%%%%%%%%%%%%%%%%%%%%%
\begin{figure}
\caption{The asymmetry energy as a function of density for the {\bf IS}
(dashed curve) and {\bf ISB} (full curve) cases.
}
\label{fig5}
\end{figure}
%%%%%%%%%%%%%%%%%%%%%%%%%%%%%%%%%%%%%%%%%%%%%%%%%%%%%%%%%%%%%%%%%%%
\begin{figure}
\caption{The total energy of $\beta$-stable matter as a function of
density for the {\bf IS} (dashed curve) and {\bf ISB} (full curve) cases.
}
\label{fig6}
\end{figure}
%%%%%%%%%%%%%%%%%%%%%%%%%%%%%%%%%%%%%%%%%%%%%%%%%%%%%%%%%%%%%%%%%%%
\begin{figure}
\caption{As Figure 6, but for the energy contribution of leptons.
}
\label{fig7}
\end{figure}
%***********************************************************************
\begin{figure}
\caption{As Figure 6, but for the proton fraction.
}
\label{fig8}
\end{figure}
%%%%%%%%%%%%%%%%%%%%%%%%%%%%%%%%%%%%%%%%%%%%%%%%%%%%%%%%%%%%%%%%%%%
\begin{figure}
\caption{As Figure 6, but for the equation of state of
$\beta$-stable matter.
}
\label{fig9}
\end{figure}
%***********************************************************************
\newpage
\begin{table}
\caption{
The comparison between our results for the energy differences of {\bf ISB}
and {\bf IS} cases at $\beta = 0.0$ and those of MPM (Muther et al., 1999).
}
\begin{center}
\begin{tabular}{|c|cccccc|}
\hline
&&&$E_{ISB}-E_{IS}(MeV)$&&&\\
\cline{2-7}
Density&Our results&MPM&MPM&MPM&MPM&MPM\\
($fm^{-3}$)&&({\it Reid-93})&({\it Nijm-I})&({\it Nijm-II})&({\it CD Bonn})&
({\it $AV_{18}$})\\
\hline
0.10&0.14&0.17&0.22&0.16&0.22&0.21\\
0.15&0.21&0.21&0.28&0.18&0.28&0.25\\
0.25&0.34&0.25&0.51&0.32&0.51&0.41\\
0.35&0.48&0.28&0.78&0.43&0.79&0.55\\
0.40&0.56&0.29&0.87&0.51&0.89&0.61\\
\hline
\end{tabular}
\end{center}
\label{tab1}
\end{table}
%***********************************************************************

\begin{table}
\caption{The effects of {\bf ISB} of nucleon-nucleon interaction on the
staturation density ($fm^{-3}$), energy (MeV) and incompressibility (MeV)
of asymmetrical nuclear matter for different
values of asymmetry parameter $\beta$.
}
\begin{center}
\begin{tabular}{|c|ccc|ccc|}
\hline
&&{\bf ISB}&&&{\bf IS}&\\
\cline{2-7}
$\beta$&Density&Energy&Incompressibility&
Density&Energy&Incompressibility\\
\hline
0.0&0.272&-13.81&224&0.277&-14.08&233\\
0.33&0.245&-11.39&196&0.252&-11.85&206\\
0.66&0.185&-3.47&112&0.196&-4.02&124\\
\hline
\end{tabular}
\end{center}
\label{tab2}
\end{table}
%***********************************************************************
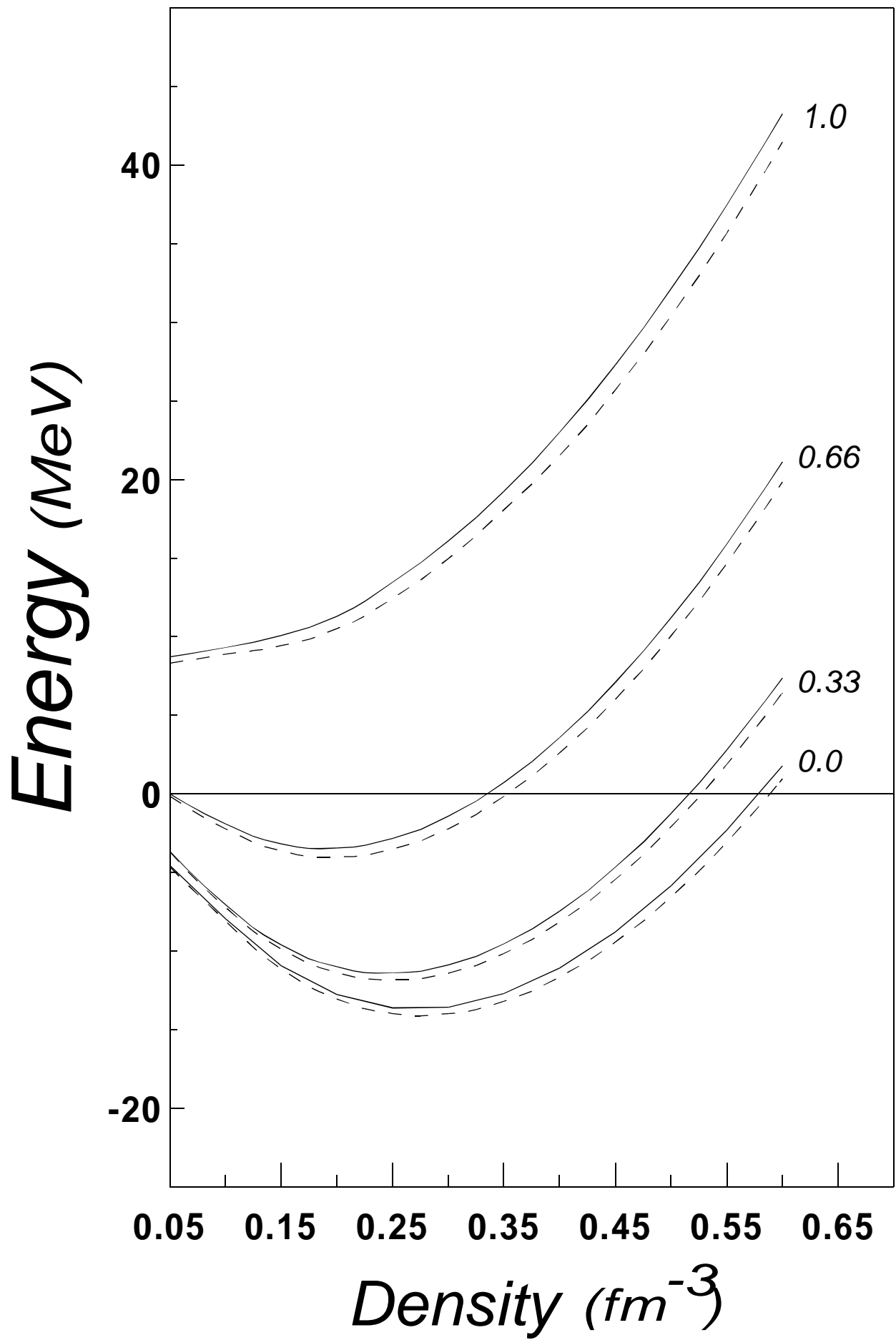
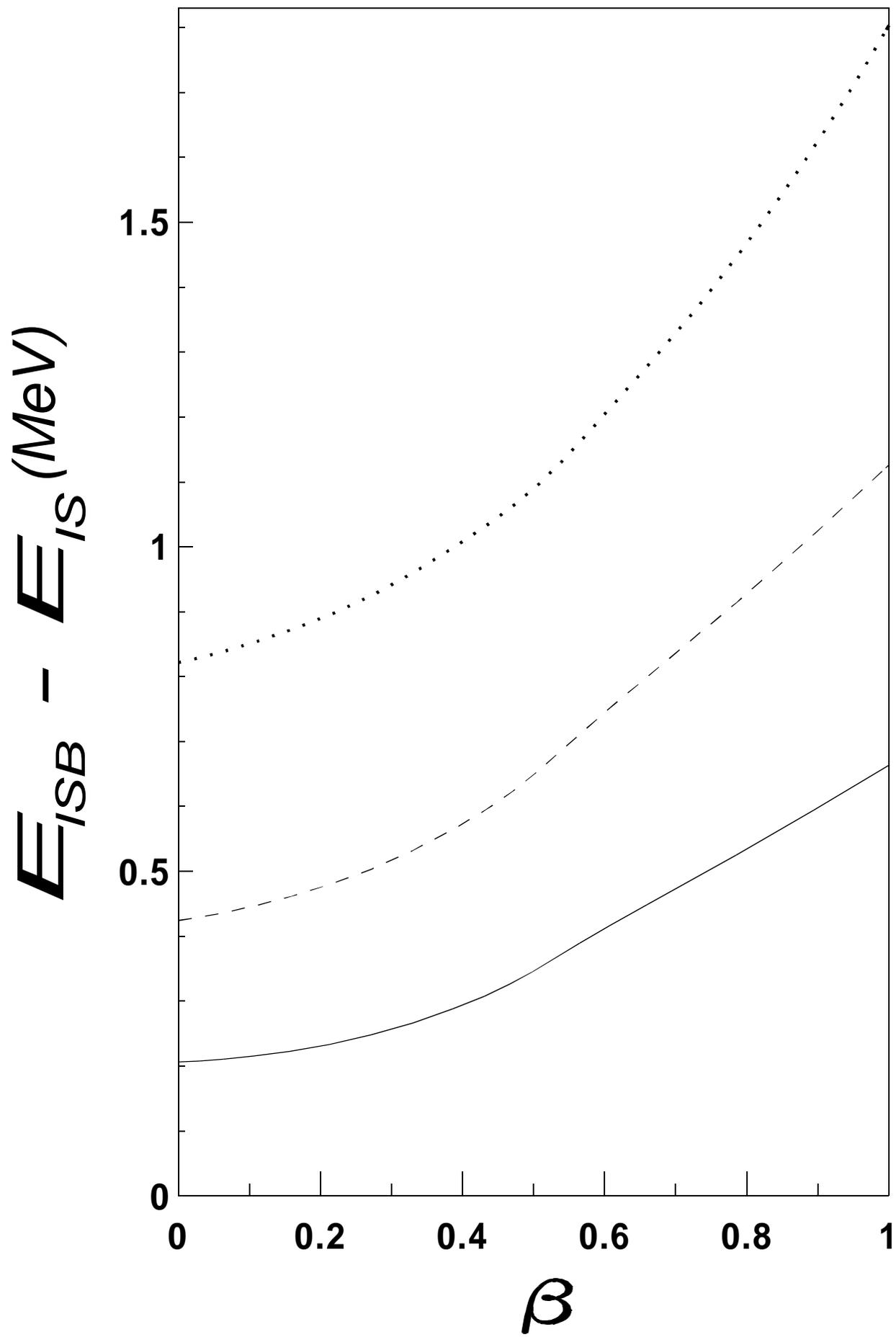
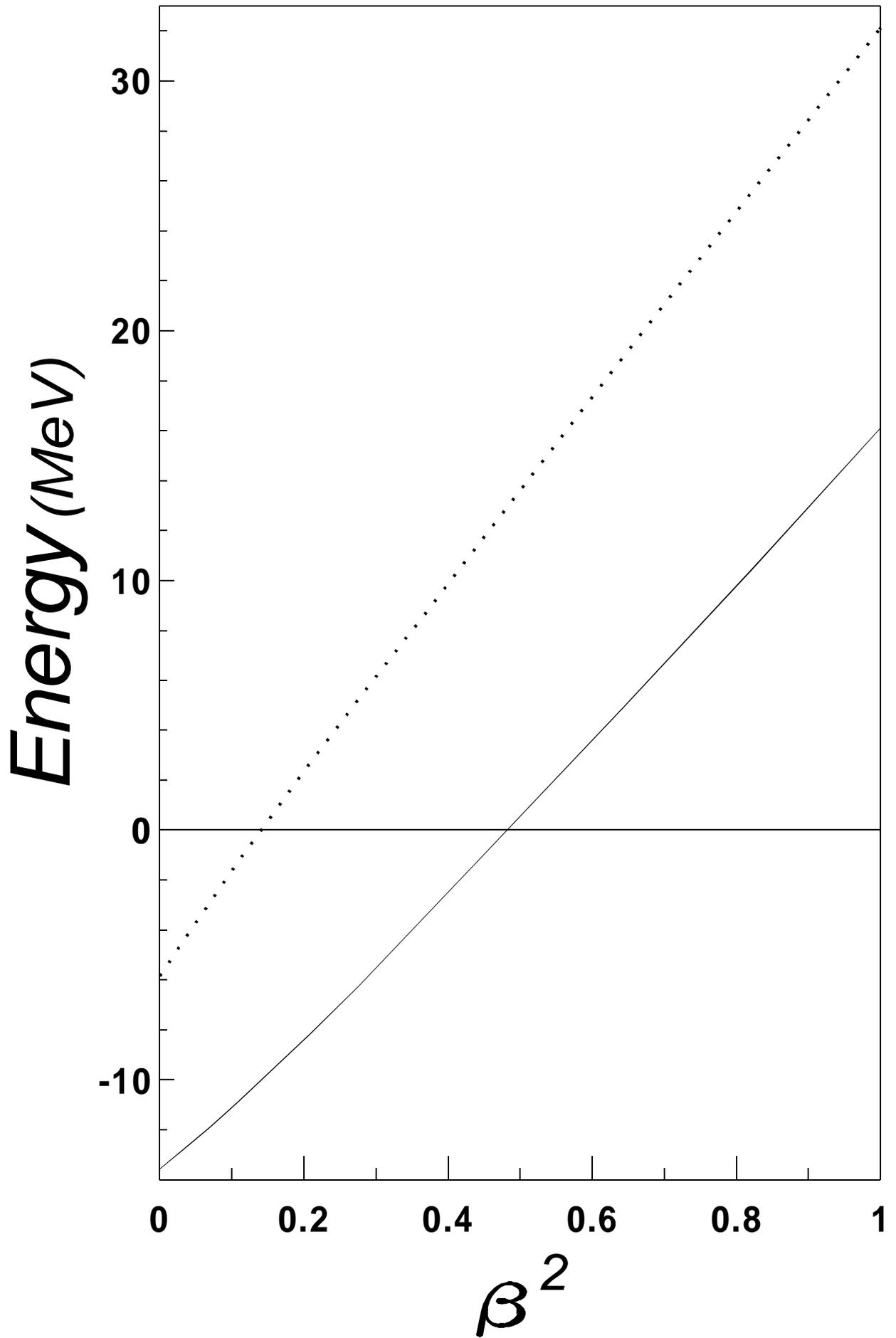
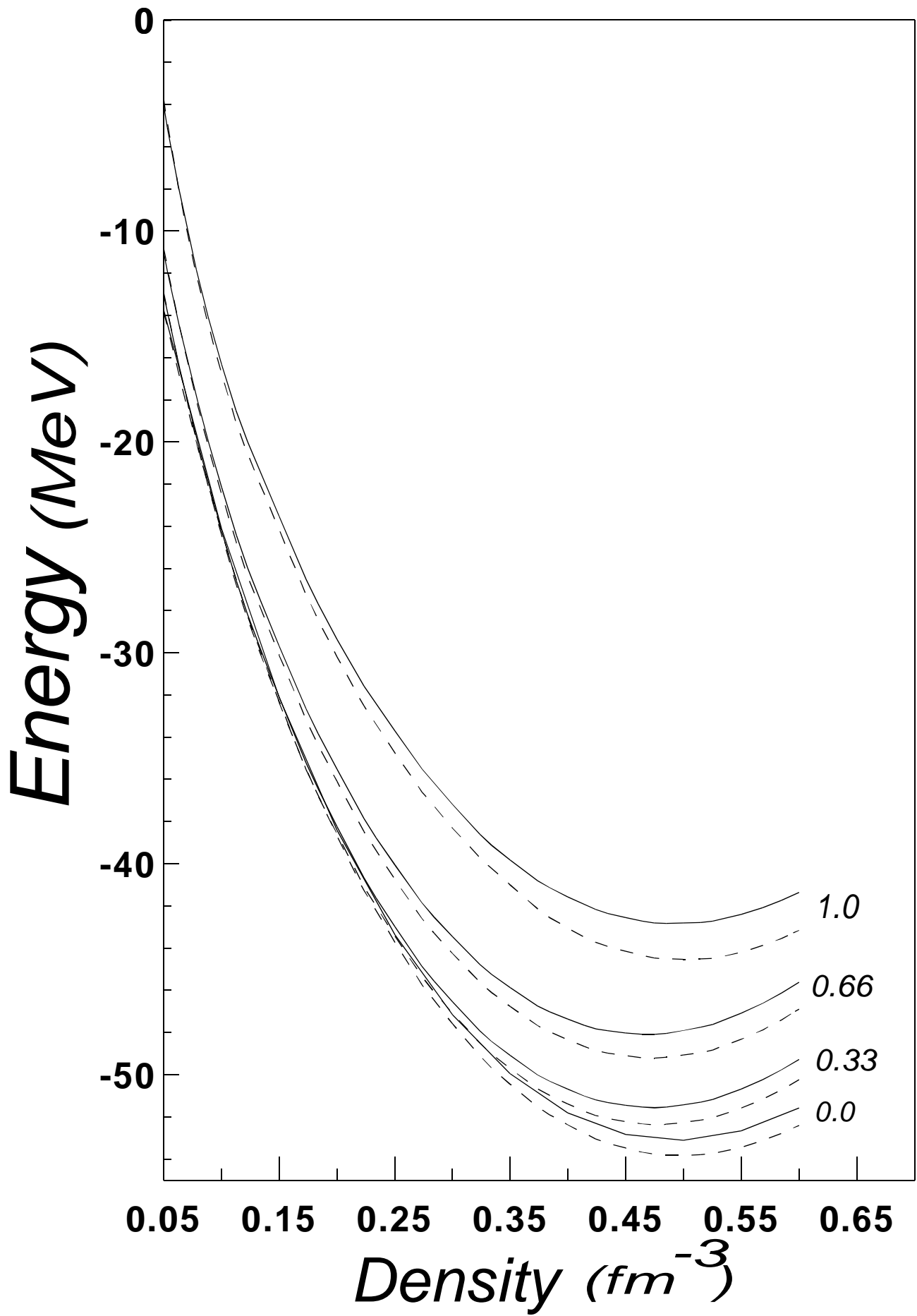
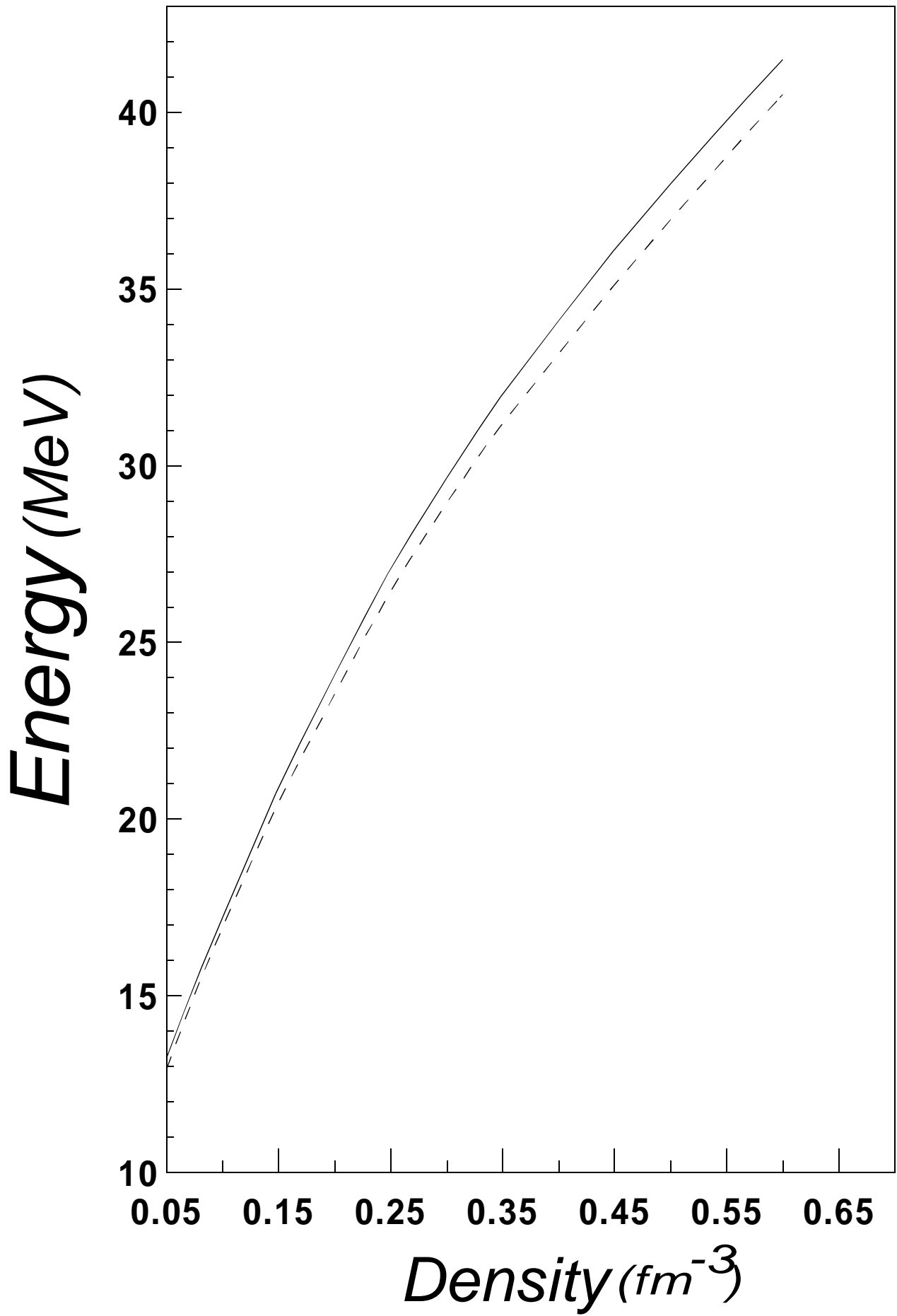
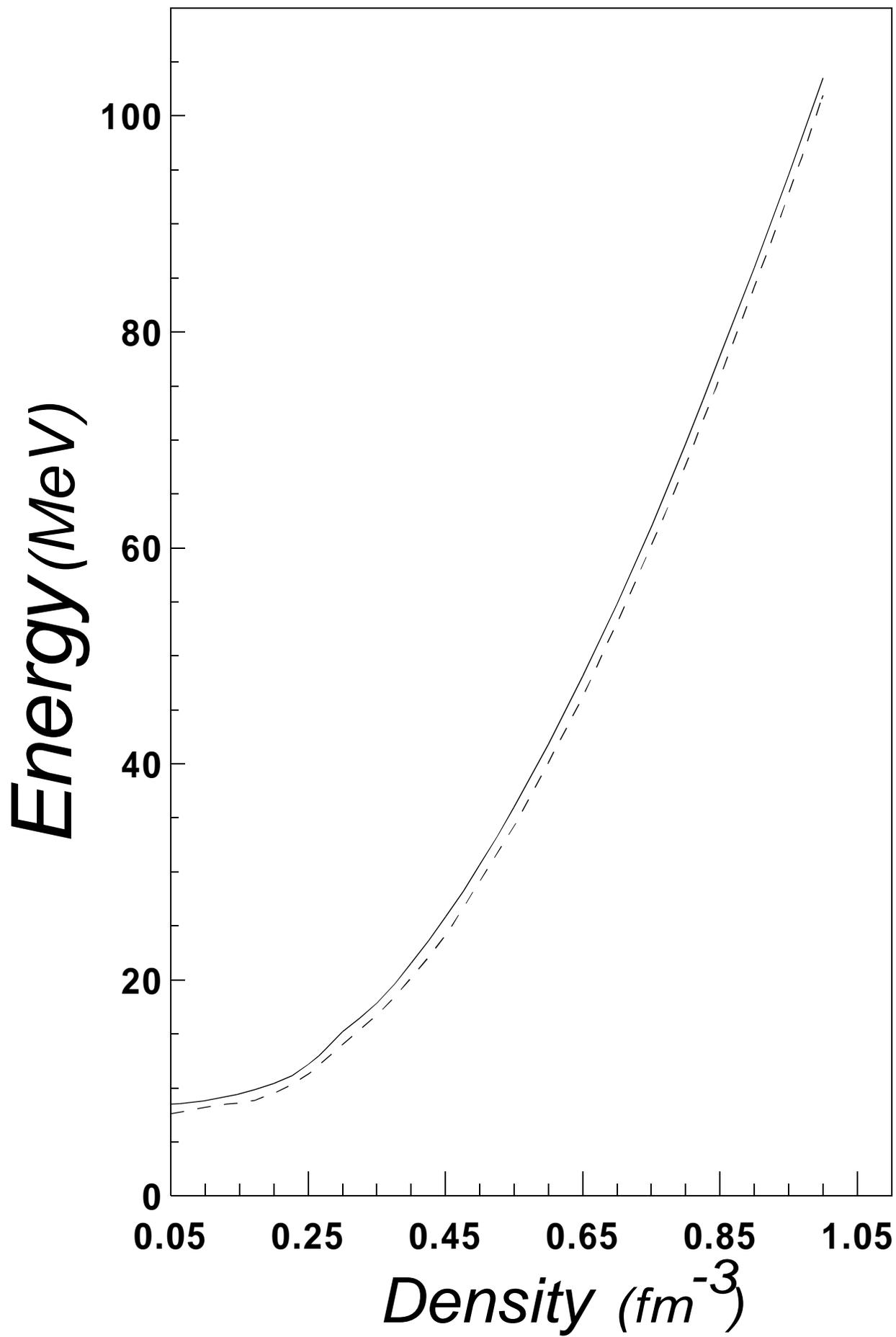
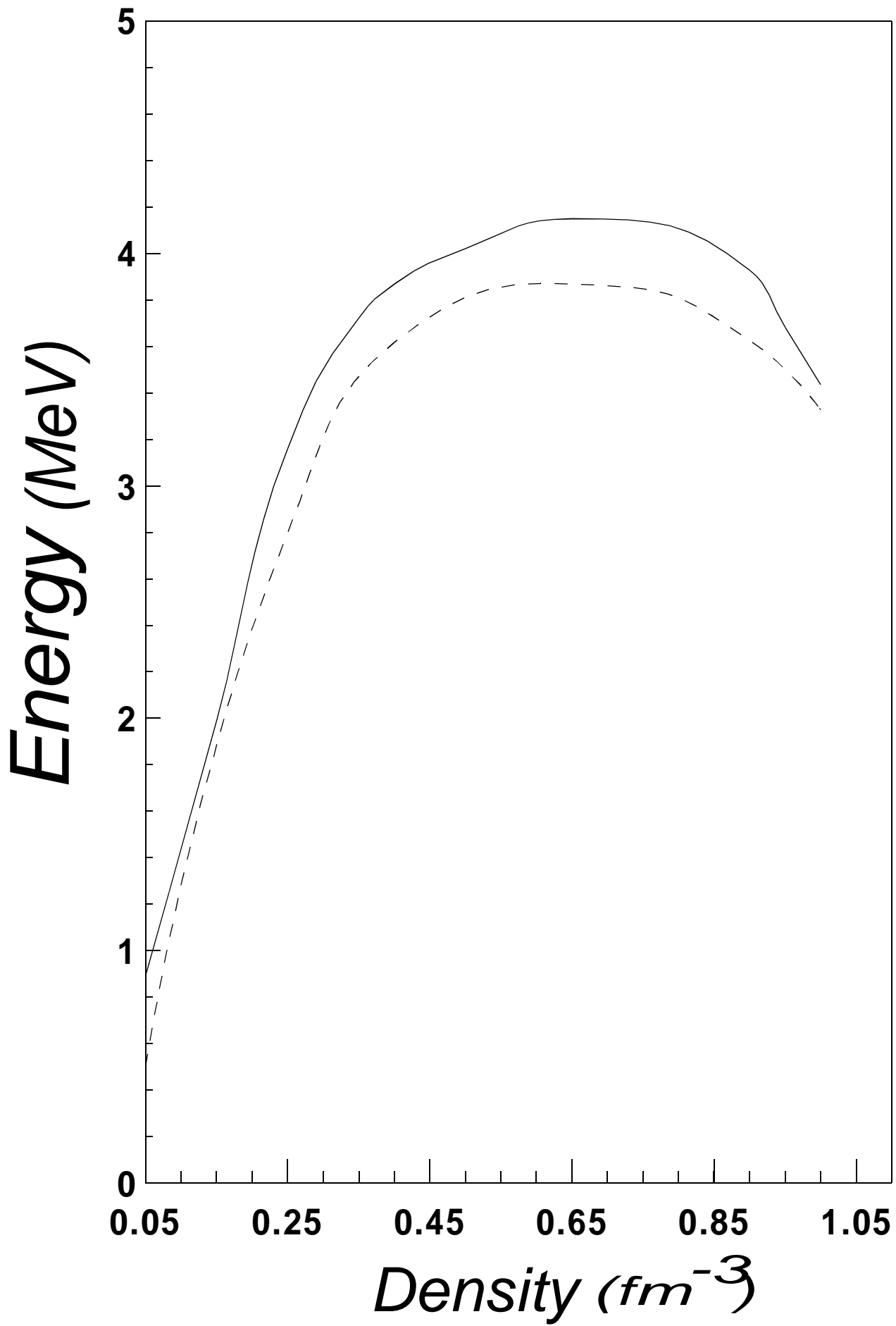
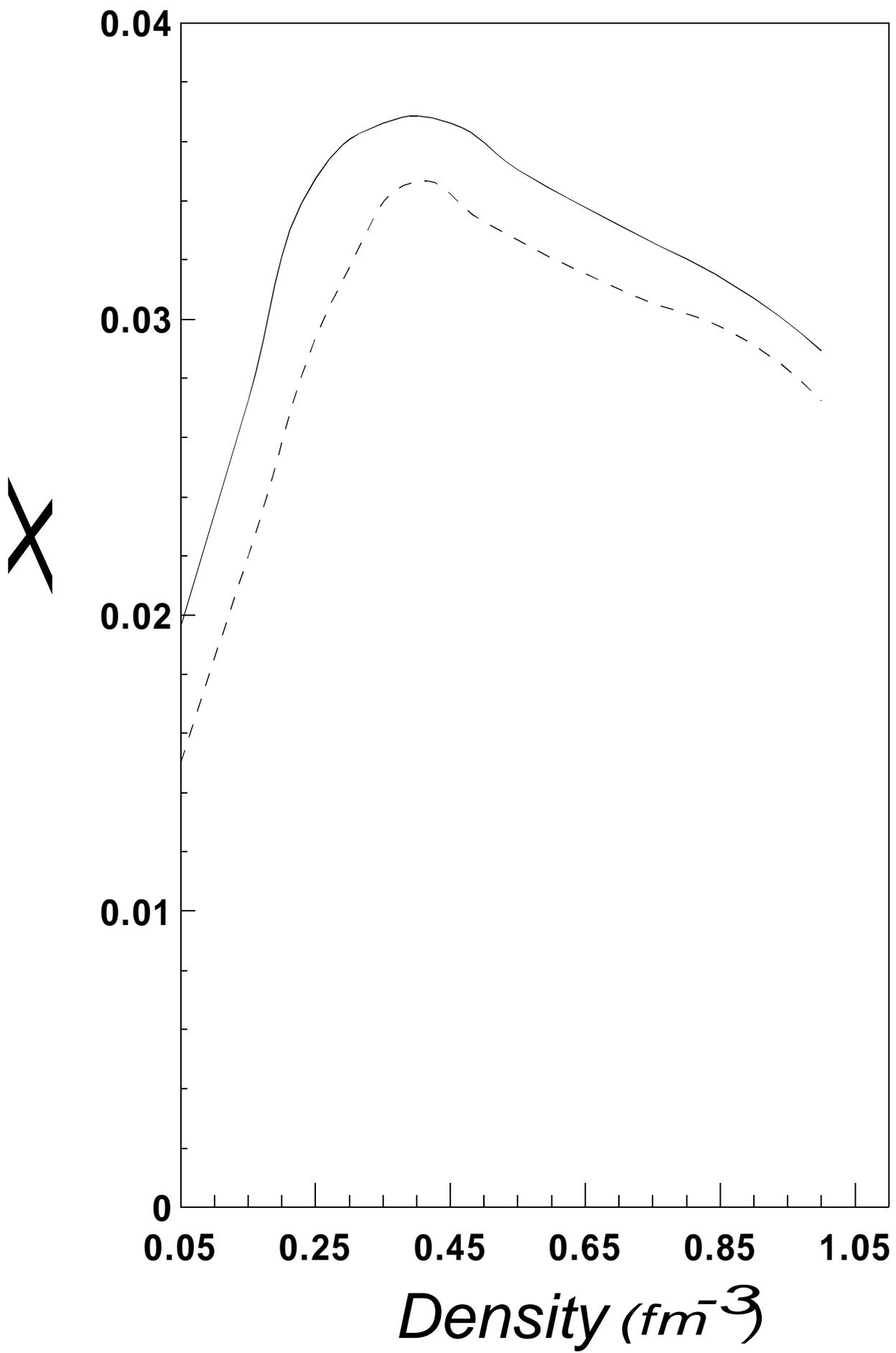
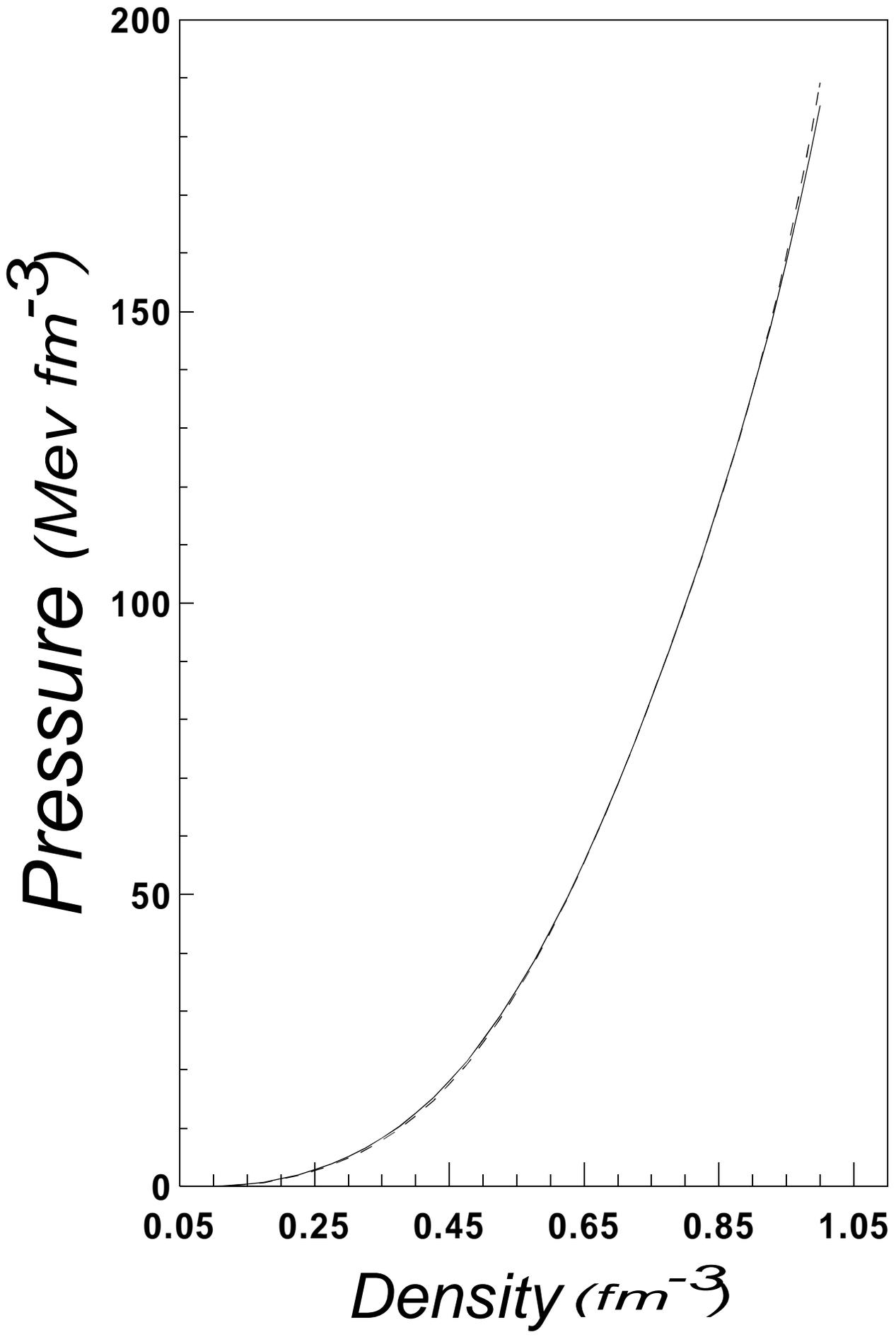
\end{document}